# Surpassing fundamental limits of oscillators using nonlinear resonators

L.G. Villanueva[1], E. Kenig[1], R.B. Karabalin[1], M.H. Matheny[1], R. Lifshitz[2], M.C. Cross[1], M.L. Roukes[1]

[1]*Kavli Nanoscience Institute, California Institute of Technology, Pasadena, CA, 91125*
[2]*Raymond and Beverly Sackler School of Physics and Astronomy, Tel Aviv University, 69978 Tel Aviv, Israel*

**ABSTRACT**

Self-sustained oscillators are ubiquitous and essential for metrology, communications, time reference, and geolocation. In its most basic form an oscillator consists of a resonator driven on-resonance, through feedback, to create a periodic signal sustained by a static energy source. The generation of a stable frequency, the basic function of oscillators, is typically achieved by increasing the amplitude of motion of the resonator while remaining within its linear, harmonic, regime. Contrary to this conventional paradigm, in this Letter we show that by operating the oscillator at special points in the resonator's anharmonic regime we can overcome fundamental limitations of oscillator performance due to thermodynamic noise as well as practical limitations due to noise from the sustaining circuit. We develop a comprehensive model that accounts for the major contributions to the phase noise of the nonlinear oscillator. Using a nanoelectromechanical system (NEMS)-based oscillator, we experimentally verify the existence of a special region in the operational parameter space that enables a significant reduction of the oscillator's phase noise, as predicted by our model.

Advances in time and frequency measurement have closely paralleled technological progress. However, since the appearance of quartz-crystal-based oscillators[1], very few conceptual innovations have been introduced: quartz crystal resonators (their frequency-determining elements) operate at the highest possible signal to noise ratio in order to minimize phase noise. The resonator is always kept within its *linear* regime, which results in oscillator phase noise being inversely proportional to the oscillator carrier power. Ongoing technological evolution requires a dramatic reduction in the oscillator size and power, preferably without performance degradation. Micro- and nano-electromechanical systems[2-5] are increasingly being considered as valid alternatives to quartz as the frequency-determining element. However, with the reduction in size, their dynamic range also diminishes since nonlinear effects manifest at lower amplitudes[6,7]. This has proven interesting for fundamental studies[8-10], but is typically considered detrimental to the oscillator performance[11,12]. However, several techniques have been proposed to utilize *nonlinear* behavior in the mechanical element in order to improve oscillator performance. These proposals rely on the local elimination of frequency to energy dependence[13], evasion of amplifier noise[14], use of either parametric feedback[15], non-degenerate parametric drive[16] or coupling to internal resonances[15,17].

In this Letter, we analyze all the contributions to the phase noise in an oscillator based on a *nonlinear* resonator. We predict the existence of a special region in the parameter space, above the nonlinear threshold, where the dominant contributions to the phase noise are suppressed. We construct such an oscillator from a nanomechanical doubly-clamped beam resonator and measure its phase noise. We find remarkable agreement with our theoretical model, and unequivocally confirm experimentally the existence of such a special region, where the phase noise performance is improved beyond the limitations of the linear regime. Our findings contravene conventional phenomenological wisdom, which assumes that operation beyond the threshold of nonlinearity necessarily degrades phase noise. Indeed by operating the oscillator in this region, the signal level can be increased to large values without the conventionally expected performance degradation. It is therefore possible to overcome fundamental limitations of oscillator performance due to thermodynamic noise.

Since we are interested in slow modulation dynamics of an oscillator constructed from a high-Q resonator, we introduce a dimensionless slow time scale $T = \varepsilon \omega_0 t$ with $\varepsilon$ a small expansion parameter chosen for convenience as detailed below. The resonator displacement is given by $x(t) = x_0 \operatorname{Re}[A(T)e^{i\omega_0 t}] + \cdots$, with $x_0$ being a convenient scale factor, Re standing for real part, and the ellipses ($\cdots$) representing negligible harmonics generated by the resonator nonlinearity. Our theoretical analysis is based on the dimensionless equation of motion for the complex amplitude $A(T) = a(T)e^{i\phi(T)}$ of the resonator dynamics:

$$\frac{dA}{dT} = -\frac{\gamma}{2}A + i\frac{3}{8}\alpha|A|^2 A + \frac{i}{2}F(a)e^{i\phi}e^{i\Delta}. \tag{1}$$

The first two terms on the right hand side of Eq. (1) arise, respectively, from the linear dissipation, and from the essential nonlinearity of the resonator, i.e. the dependence of the resonance frequency on the amplitude of motion. The third term represents the feedback loop drive projected onto the slow equation of motion of the resonator. The behavior of the feedback loop is then described by the gain function $F(a)$ and the phase delay $\Delta$ relative to the resonator phase. Equation (1) relies on the assumption of *weak* feedback (just sufficient to overcome the small dissipation of the high-Q resonator); then the amplitude of the motion is *small*, so that nonlinear frequency shifts are comparable to the linear resonance line width, but small compared to the resonance frequency $\omega_0$.

Equation (1) is derived from the basic equation of motion using secular perturbation theory[18], and our results are generally applicable. However, to make the discussion concrete we will focus on our particular experimental demonstration, based on a NEMS resonator. The parameters $\gamma$ and $\alpha$ are related to the quality factor $Q$ and to the nonlinear coefficient $\tilde{\alpha}$ in the spring constant $m\omega_0^2(1 + \tilde{\alpha}x^2)$ according to:

$$\gamma = \frac{1}{Q\varepsilon}, \qquad \alpha = \frac{\tilde{\alpha}x_0^2}{m\omega_0^2 \varepsilon}, \tag{2}$$

with $m$ the mass of the resonator. For the perturbation theory to be consistent $\gamma, \alpha$ must be $O(1)$ quantities. Thus we choose scale factors $\varepsilon = Q^{-1}$ and $x_0^2 = m\omega_0^2/\tilde{\alpha}Q$ so that in the absence of fluctuations $\gamma$ and $\alpha$ are unity.

We focus our study on a heavily-saturated oscillator, that is, one in which the system gain is designed to keep the feedback magnitude constant regardless of the amplitude of motion. This scheme is also known as phase feedback oscillator[14,19], which is commonly used to suppress one quadrature of the amplifier noise. It also provides, in principle, a quantum nondemolition[20] method to track the resonator phase. For saturated feedback, Eq. (1) reduces to

$$\frac{dA}{dT} = -\frac{\gamma}{2}A + i\frac{3}{8}\alpha|A|^2 A - i\frac{s}{2}e^{i\phi}e^{i\Delta} \tag{3}$$

with $s$ the saturation level. This equation can be separated into equations for the magnitude $a$ and phase $\phi$

$$\begin{aligned}\frac{da}{dT} &= -\frac{\gamma a}{2} + \frac{s}{2}\sin\Delta \equiv f_a; \\ \frac{d\phi}{dT} &= \frac{3\alpha}{8}a^2 - \frac{s}{2}\frac{\cos\Delta}{a} \equiv f_\phi.\end{aligned} \tag{4}$$

For steady state oscillations $da/dT = 0$, $d\phi/dT = \Omega$, with $\Omega$ giving the frequency offset of the oscillations from the linear resonance frequency, in units of the resonator line width. Thus, Eqs. (4) yield expressions for the oscillation amplitude and frequency offset that define the limit cycle:

$$a = \frac{s}{\gamma}\sin\Delta, \quad \Omega = \frac{d\phi}{dT} = \frac{3\alpha}{8}a^2 - \frac{s}{2}\frac{\cos\Delta}{a}. \tag{5}$$

Our experimental demonstration is performed using a piezoelectric NEMS doubly-clamped beam made from an aluminum nitride (AlN) and molybdenum (Mo) multilayer (Figure 1). In our experimental implementation (Figure 1a), both the phase delay, $\Delta$, and the power of the feedback, $s$, can be externally and independently controlled. This permits full exploration of the input parameter space of the feedback oscillator. We first confirm that the system behaves according to predictions for a heavily-saturated oscillator (Figure 1b). For periodic solutions $\phi = \phi_0 + \Omega T$ the equation of motion (3) for the heavily-saturated oscillator is identical to the one for an open-loop resonator externally driven with a periodic source of constant magnitude $s$; then $\Delta$ represents the phase difference between the resonator motion and the drive. As is known for nonlinear resonators, when the driving force is sufficiently large, the system can bifurcate into three possible solutions at a given drive frequency: two of these are stable, and one is unstable[18]. In the case of the heavily saturated oscillator, the system also presents three possible values for the amplitude of oscillation at a given frequency above a threshold feedback power. However, in this latter case, the resonator-drive phase difference is itself determined by the feedback, and both amplitude and frequency are single-valued functions of this phase. Therefore, all three operating conditions at the same frequency might be stable[19], and this is indeed confirmed by a stability analysis using Eq. (3), and our measurements (Figure 1.b).

We now turn to the noise analysis of the feedback-sustained oscillator. In general, the noise, when projected onto the slow dynamics, is represented by adding a complex stochastic term $\Xi_R(T) + i\Xi_I(T)$ to the evolution in Eq. (1). The performance of an oscillator is typically characterized by the spectral density of its phase noise, $S_\phi$, or the variance $\langle[\delta\phi(T+\tau) - \delta\phi(T)]^2\rangle$ of the phase deviation $\delta\phi(T) = \phi(T) - \Omega T$, which can be found by solving Eq. (1) with the additional stochastic terms.

For our saturated feedback NEMS oscillator it is possible to distinguish two types of noise affecting the phase diffusion of the oscillator: thermomechanical noise and parameter noise[20,21]. Thermomechanical noise is caused by the Brownian motion of the resonator: it enters the equation as a random, perturbative force and affects independently both quadratures of the oscillation with the same intensity. Its projection in quadrature to the displacement (the phase direction) always affects the oscillator performance (hereafter called the *direct thermomechanical contribution*), whereas its projection in the amplitude direction affects the phase noise only through *amplitude-phase conversion*. This is typically assumed to be dominant at higher amplitudes when nonlinear resonators are used. Parameter noise is caused by fluctuations in the parameters $p_i$ determining the oscillator operational point (in our case $\gamma$, $s$, $\alpha$ and $\Delta$). Each independent noise source, $n$, is described by stochastic terms $v_{a,n}\Xi_n(T)$, $v_{\phi,n}\Xi_n(T)$ added to the amplitude and phase evolution equations (4) respectively, where the noise vector $(v_{a,n}, v_{\phi,n})$ gives the relative strength of the $n$th noise force in the amplitude and phase quadratures.

Two key points lead to our predictions for reducing the frequency precision degradation. Firstly, for small frequency offsets compared to the amplitude relaxation rate (i.e. the resonator line width) the time derivative term $da/dT$ can be neglected in calculating the amplitude fluctuations. Secondly, the evolution terms $f_a, f_\phi$ in Eqs. (4) do *not* depend on the phase $\phi$: this is the basic phase symmetry of the limit cycle when Eq. (1) applies. These lead directly to a long-time phase diffusion, which is given for each independent noise source by:

$$\frac{d\delta\phi}{dT} = \sqrt{D_n}\Xi_n(T) \tag{6}$$

with

$$D_n = \left(v_{\phi,n} - \frac{\partial f_\phi/\partial a}{\partial f_a/\partial a} v_{a,n}\right)^2. \tag{7}$$

We model each noise term $\Xi_n(T)$ as white noise of intensity $I_n$, $\langle \Xi_n(T)\Xi_n(T')\rangle = I_n\delta(T-T')$. The phase variance is then found, by integration and averaging of Eq. (6), to grow linearly in time (diffusively), with a diffusion constant $D_n I_n$, i.e. $\langle[\delta\phi(T+\tau) - \delta\phi(T)]^2\rangle = (\sum_n D_n I_n)\tau$. These results can be formally derived using a spectral analysis of the stochastic fluctuations, or following the methods of Demir et al.[21] for phase diffusion of a general limit cycle. One can generalize these results to other noise spectra, such as $1/f$.[22]

The first term in Eq. (7) represents the direct effect of the $n$th noise source on the oscillator phase; the second term accounts for phase diffusion due to amplitude-phase conversion. Furthermore, for noise due to fluctuations in the parameter $p_i$, the noise vector becomes: $v_{\phi,i} = \partial f_\phi/\partial p_i$, $v_{a,i} = \partial f_a/\partial p_i$, and the expression for $D_n$ reduces to

$$D_n = D_{p_i} = \left(\frac{d\Omega}{dp_i}\right)^2, \tag{8}$$

so that the stochastic phase diffusion can be evaluated immediately from the dependence of the oscillator frequency on the parameters. Alternatively, for thermomechanical noise that is purely in the amplitude quadrature ($v_a = 1$, $v_\phi = 0$), the coefficient which quantifies the strength of amplitude-phase conversion is

$$D_a = \left(\frac{\partial \Omega}{\partial a} \Big/ \frac{\partial f_a}{\partial a}\right)^2 = 4\left(\frac{\partial \Omega}{\partial a}\right)^2; \qquad (9)$$

whereas for thermomechanical noise that is purely in the phase quadrature ($v_a = 0$, $v_\phi = 1/a$) the strength of direct thermomechanical noise contribution to the phase noise is $D_{direct} = 1/a^2$.

Combining the above results, the total phase noise as a function of the offset frequency $\delta f$ is given by the sum

$$S_\phi(\delta f) = \frac{1}{2\pi} \frac{f_c}{Q} \frac{1}{\delta f^2} \sum_n I_n D_n; \qquad (10)$$

where $f_c$ is the carrier frequency and the parameters $D_n$ have been defined above and the expressions are expanded in Table 1. Note that the dependence on $\delta f^{-2}$ emerges from the assumption of the noise terms being white. As we show elsewhere[22], a similar result is obtained if colored noise is considered.

Equation (10) shows two strategies for oscillator performance optimization: minimization of either $I_n$ or $D_n$. In this Letter, we focus on the latter – both for its general applicability and because the $D_n$ terms are experimentally controllable parameters, whereas the $I_n$ coefficients are dictated by the environment. Further, we pay special attention to the terms that are typically considered to be dominant: $D_{direct}$, $D_a$ and $D_\Delta$.

The direct contribution of thermomechanical noise has been widely analyzed in the literature[23] and is suppressed by maximizing the oscillator amplitude ($a$). Noise in the feedback phase ($\Delta$) can be cancelled at the operational points where $D_\Delta = (d\Omega/d\Delta)^2 = 0$. Greywall and Yurke[14,24] proposed the operation at the bifurcation point, where this condition is satisfied, and showed that near such a "*Duffing critical point*" (DCP) the oscillator's phase is unaffected by fluctuations in $\Delta$. We extend this understanding further and note that above the threshold of nonlinearity, for each saturation value, there are actually *two* values of Δ for which $d\Omega/d\Delta = 0$. At the bifurcation ($s = s_c$), the case considered by Greywall and Yurke[14,24], both of these DCPs are degenerate at $\Delta = 120°$. However, for larger feedback powers, one family of DCPs approaches $\Delta = 90°$ while the other one tends toward $\Delta = 180°$ (see Supplementary Information, Figure S2).

From Eq. (9) we conclude that amplitude-phase converted thermal noise can be cancelled at the points where $\partial\Omega/\partial a = 0$ (note that this is not where the total derivative vanishes, i.e. $d\Omega/da = 0$)[25]. This term has always been considered to be zero when the resonator used is linear and assumed to be unavoidable when the resonator used is nonlinear. In fact, we show that for linear resonators this is only true for a particular feedback phase, $\Delta = 90°$, and, equivalently, for nonlinear resonators there also exists a value of Δ for each feedback power such that $\partial\Omega/\partial a = 0$, effectively detaching amplitude and phase. We call this the "*amplitude detachment point*" (ADP). Importantly, according to our model, the location of the ADP turns out to be very close to the second aforementioned family of DCPs (see Supplementary Information, Figure S2), therefore this yields a *region* where two of the major contributions to phase noise can be drastically reduced.

In order to experimentally verify the predicted behavior, we measure the phase noise of the heavily-saturated oscillator from Figure 1 for different values of the feedback power, $s$, and phase, Δ. Figure 2 shows the results obtained at $\delta f = 1\text{kHz}$ offset from the carrier (colored spheres). Solid-black lines correspond to the predictions of the model described above (and in the Supplementary Information). In order to perform such a quantitative comparison, we first independently estimate $I_{Th}$ and $I_\Delta$, and subsequently find a bound for the value of $I_S$. We then adjust the values for all three intensities to get the best possible match (see Supplementary Information, Section C). The agreement between the experiments and theory is remarkable. Figure 2 indicates that if the resonator is operated above its onset of nonlinearity ($s = s_c = 1.433$) the phase noise near the conventional operational point ($\Delta = 90°$) is indeed increased. However, when operating near the second set of DCPs and close to the ADP, a significant performance improvement beyond what is possible in the linear regime can be achieved.

Using our model, we also gain insight into the decomposition of the observed phase noise according to the physical origins of the fluctuations, as can be seen in Figure 2. We show that thermomechanical noise and noise in $\Delta$ are the dominant contributions for most values of the phase. For high saturation values, however, it can also be observed that the phase noise at the minimum is not dominated by either of those contributions. There is a different component that is only visible around that region (and is hidden otherwise) which corresponds to fluctuations in the resonant frequency of the mechanical resonator, possibly arising from environmental noise[26,27]. The diffusion coefficient for this parameter is constant over all of the parameter space, hence this component of oscillator noise cannot be reduced by tuning the feedback phase. This specific parameter fluctuation imposes a bound on the phase noise reduction that is achievable with this NEMS device (see Supplementary Information, Section D). However, even with this ultimate limitation the phase noise is rendered significantly lower than is possible using conventional "linear" schemes.

In summary, we theoretically predict and experimentally demonstrate a fundamental and simple oscillator paradigm that harnesses nonlinear stiffness, in which the phase noise is substantially lower than in "linear" operation. At the newly-identified special points in the $s - \Delta$ parameter space, the effects of fluctuations in the feedback phase are eliminated and *amplitude-phase conversion* of the thermomechanical noise is suppressed. This optimization contravenes conventional wisdom and establishes a new cornerstone for the use of nonlinear resonators as a frequency determining element in self-sustained oscillators. We highlight that these results are applicable not only to NEMS as used here, but can also be used for *any* type of resonator (electrical, optical, etc.) that possesses nonlinearity[28-31].

| Type of noise | | Diffusion susceptibility to a parameter | Noise intensity |
|---|---|---|---|
| Thermo-mechanical | Direct | $D_{Th,direct} = \dfrac{1}{a^2}$ | $I_{Th}$ |
| Thermo-mechanical | $A - \phi$ conversion | $D_{Th,A-\phi\ conv} = \left(\dfrac{\partial \Omega_o}{\partial a}\right)^2 \Big/ \left(\dfrac{\partial f_a}{\partial a}\right)^2 = 4\left(\dfrac{3}{4}a + \dfrac{s}{2a^2}\cos\Delta\right)^2$ | $I_{Th}$ |
| Parameter noise - $\Delta$ | | $D_\Delta = \left(\dfrac{d\Omega_o}{d\Delta}\right)^2 = \left(\dfrac{1}{2\sin^2\Delta} + \dfrac{3s^2 \sin\Delta\cos\Delta}{4}\right)^2$ | $I_\Delta$ |
| Parameter noise - $s$ | | $D_s = \left(\dfrac{d\Omega_o}{ds}\right)^2 = \left(\dfrac{3}{4}\dfrac{a^2}{s}\right)^2$ | $I_s$ |
| Parameter noise - $\alpha$ | | $D_\alpha = \left(\dfrac{d\Omega_o}{d\alpha}\right)^2 = \left(\dfrac{3}{8}a^2\right)^2$ | $I_\alpha$ |
| Parameter noise - $\gamma$ | | $D_\gamma = \left(\dfrac{d\Omega_o}{d\gamma}\right)^2 = \left(a\left(\dfrac{3a}{4} + \dfrac{s}{2a^2}\cos\Delta\right)\right)^2$ | $I_\gamma$ |
| Parameter noise – $\omega_0$ | | $D_{\omega_0} = \dfrac{1}{4}$ | $I_{\omega_0}$ |

**Table 1** – Diffusion coefficients for different physical mechanisms affecting phase noise

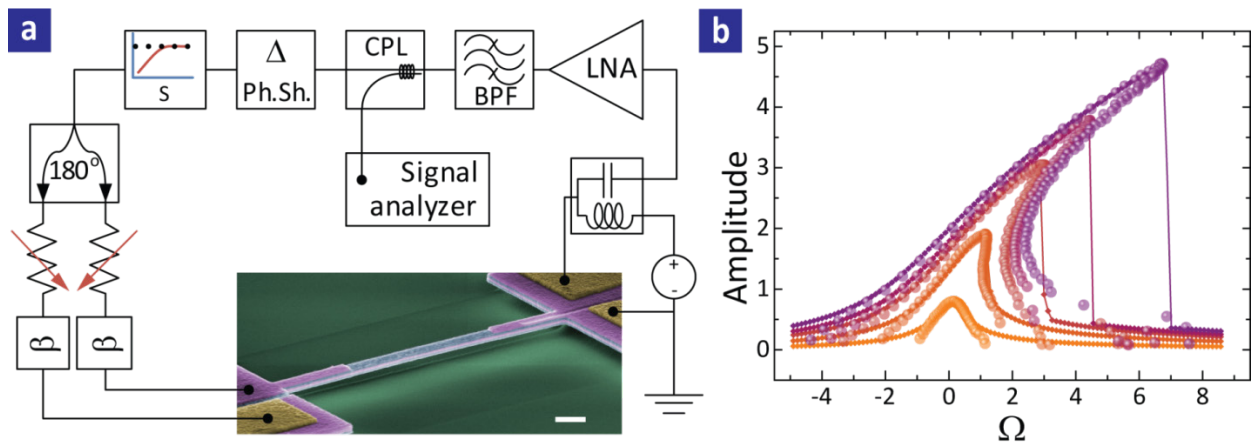

**Figure 1 | Experimental implementation.** (a) Schematic diagram of the feedback system. The signal from the resonator is amplified and filtered to eliminate higher harmonics and noise. After an externally controlled phase delay (Δ) is applied, the signal is passed through a variable limiter to select the power fed back to the resonator through a background cancellation differential bridge. The bridge is required to avoid undesirable RF cross-talk that is not due to mechanical motion. Colored SEM micrograph shows the device utilized for the experiments in this paper: a doubly-clamped beam composed of a four-layer stack of AlN(20nm)/Mo(100nm)/AlN(50nm) /Mo(50nm), a width of 420 nm and a length of 9 μm. The resonance frequency of the device is $f_0 = 12.63\ MHz$ and its quality factor is $Q = 1600$, both measured at room temperature and 1 mtorr. The transduction of the motion is performed via piezoelectric actuation and piezometallic (metallic gauge effect) detection, as described elsewhere[7]. The scale bar is 500 nm. (b) (Squares and Solid lines) Resonant response of the open-loop (driven) resonator for 5 different driving powers ($s = 0.5, 1.22, 3.06, 4.73, 7.23$). The vertical drops in the response correspond to the points where stability is lost. (Spheres) Oscillation amplitude vs. oscillation frequency for the heavily saturated closed-loop system, taken at five different values of $s$, which correspond to the open-loop drive levels, while sweeping the phase Δ. As predicted by theory, both responses overlap where the open-loop response is stable. Using the closed-loop system, access to otherwise unstable operation points is possible. Plotted magnitudes are dimensionless, scaling coefficients are reported in Supplementary Information.

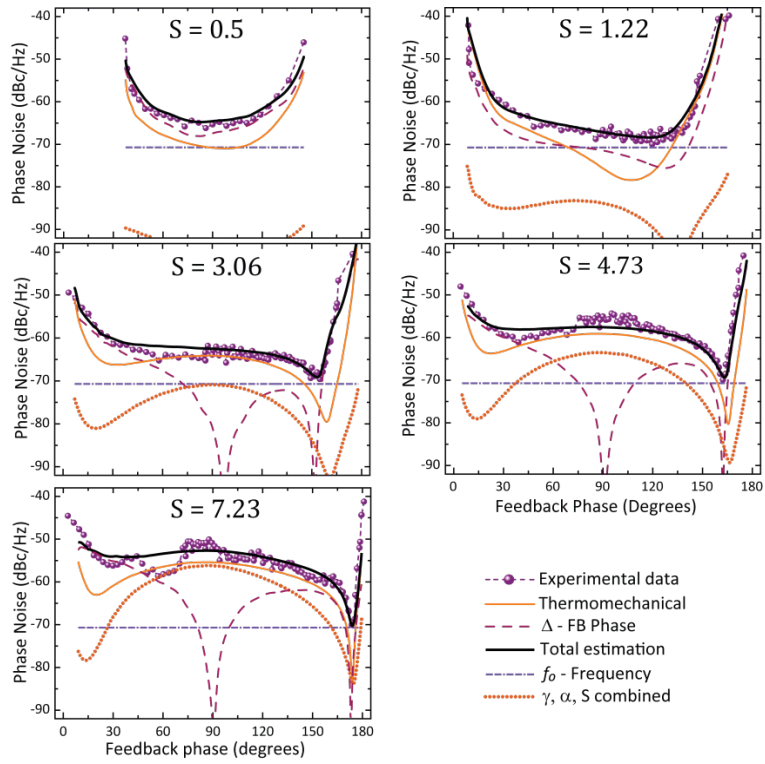

**Figure 2 | Experimental results and decomposition of phase noise according to physical mechanism.** Experimental phase noise at 1 kHz offset from the carrier ($\mathfrak{L}(1\,\mathrm{kHz}) = 10\log_{10} S_\phi(1\,\mathrm{kHz})$) for different saturation power levels is shown (spheres, dashed line), superimposed on the total theoretical estimate (black line). The calculated contributions to the total phase noise from the different sources are also shown. Thermomechanical noise (direct and amplitude-phase converted contributions plotted together) and noise in $\Delta$ dominate most of the phase range except in the region close to the amplitude-phase detachment point, where fluctuations in frequency become apparent. Fluctuations in saturation ($s$), dissipation ($\gamma$) and nonlinearity ($\alpha$) are also considered, but are plotted as their joint contribution for the sake of simplicity. Noise intensities are independently estimated in the case of thermomechanical noise and fluctuations in $\Delta$. The rest of noise intensities are adjusted to provide a better fit to experimental data, taking into account the independently estimated limitation for fluctuations in the saturation. For low saturation values ($s = 0.5$), the behavior of the phase noise is almost symmetric with respect to $\Delta = 90°$. As the saturation increases, however, the phase noise loses this symmetry and a minimum starts to appear for higher $\Delta$ values. This is more apparent beyond the critical saturation value ($s \geq s_c = 1.433$): the higher the saturation, the clearer the location of the minimum and the closer to $\Delta = 180°$. This minimum corresponds to the simultaneous minimization of noise in $\Delta$ and *amplitude-phase* conversion.